\documentclass[reprint, amsmath,amssymb, aps,floatfix,superscriptaddress]{revtex4-2}
\usepackage{lipsum}
\usepackage{float}
\usepackage{amsmath}
\usepackage{graphicx}
\usepackage{dcolumn}
\usepackage{bm}
\usepackage{xcolor}
\usepackage{hyperref}
\begin{document}

\preprint{pra}

\title{Gate tunable light-matter interaction in natural biaxial hyperbolic van der Waals heterostructures} 

\author{Aneesh Bapat}
\author{Saurabh Dixit}
\author{Yashika Gupta}
\affiliation{
Laboratory of Optics of Quantum Materials, Physics Department, IIT Bombay, Mumbai - 400076\\
}
\author{Tony Low}
\affiliation{Department of Electrical and Computer Engineering, University of Minnesota, Minneapolis, Minnesota 55455, USA}
\author{Anshuman Kumar}
 \email{anshuman.kumar@iitb.ac.in}
\affiliation{
Laboratory of Optics of Quantum Materials, Physics Department, IIT Bombay, Mumbai - 400076\\
}

\date{\today}

\begin{abstract}
The recent discovery of natural biaxial hyperbolicity in van der Waals crystals, such as $\alpha$-MoO\textsubscript{3}, has opened up new avenues for mid-IR nanophotonics due to their deep subwavelength phonon-polaritons. However, a significant challenge is the lack of active tunability of these hyperbolic phonon polaritons. In this work, we investigate heterostructures of graphene and $\alpha$-MoO\textsubscript{3} for actively tunable hybrid plasmon phonon polariton modes via electrostatic gating in the mid-infrared spectral region. We observe a unique propagation direction dependent hybridization of graphene plasmon polaritons with hyperbolic phonon polaritons for experimentally feasible values of graphene chemical potential. \textcolor{black}{We further report an application to tunable valley quantum interference in this system with a broad operational bandwidth due to the formation of these hybrid modes.} This work presents a lithography-free alternative for actively tunable, anisotropic spontaneous emission enhancement using a sub-wavelength thick naturally biaxial hyperbolic materials.
\end{abstract}

\maketitle
\section*{\label{intro}Introduction}
Two-dimensional materials\cite{Low2016,Basov2016} have enabled new avenues for quantum photonics\cite{ReserbatPlantey2021} due to their potential for strong confinement of free space radiation at the deep sub-wavelength scale. The extreme confinement of electromagnetic waves and consequently strong light-matter interaction arises from polaritons\cite{Low2016, Basov2016}-- quasiparticles formed by the hybridization of a photon with charged dipolar excitations such as plasmon (for plasmon polaritons (PPs)) or lattice vibrations (for phonon polaritons (PhPs)).  Extensive research on van der Waals (vdW) crystals\cite{Basov2016,Ma2018,TaboadaGutirrez2020}, supporting PhPs and PPs, have demonstrated inherent hyperbolic anisotropy, {that is, the permittivity being negative in one direction and positive in the other,} and their applications in various nanophotonic devices such as infrared optical components\cite{dixit2021mid,Sahoo2021,Yang2017}, thermal emitters\cite{Shiue2019}, enhanced spontaneous emission rate (SER)\cite{Andersen2010}, and others\cite{Liu2019}. In particular, anisotropic spontaneous emission rate\cite{Chen2017,Gu2012}, associated with Purcell effect\cite{Purcell1946}, has great relevance for applications to light sources\cite{Andersen2010,Lee2014}, quantum interference\cite{PhysRevB.102.045416,Gu2012}, Kerr non-linearity enhancement\cite{Chen2015}, selective coupling to dark and bright excitons and others\cite{Park2017,2108.10680}.\par  

\textcolor{black}{Traditionally, hyperbolic anisotropy is achieved by complex lithographic techniques where carefully engineered sub-wavelength nanostructure of metal/dielectric materials acts like an elemental unit to attain hyperbolic dielectric permittivity tensor\cite{Guo2020}. However, this approach relies on effective medium theory where typically close to resonances, the interaction between neighbouring units cells becomes important leading to a non-local response\cite{Zhou2009}. Moreover, the maximum possible wave-vector in these systems is limited by the size of the constituent unit cell, which is limited by fabrication constraints\cite{Gjerding2017}. Lastly, artificial HMMs mostly rely on plasmonic constituent materials for the unit cells, which inevitably leads to high Ohmic losses\cite{Gjerding2017,Low2016}. Such problems can be easily overcome via natural hyperbolic materials. Hence naturally hyperbolic vdW materials are an excellent alternative to artificially engineered sytems. However, the tunability and anisotropy of these polaritons, particularly, PhPs over large operational bandwidth in conventional vdW materials\cite{Basov2016} are major limitations for applications.}\par

Graphene has been shown to be an excellent platform for tunable SPPs via electrostatic gating over a broad spectral region\cite{Chen2012,PhysRevB.80.245435}, however, these SPPs are associated with optical losses\cite{Low2014}. On the contrary, hyperbolic PhPs are relatively long lived but are restricted within a narrow spectral bandwidth\cite{Hu2019}. To this end, the integration of graphene with h-BN \cite{Woessner2014,Kumar2015,Maia2019} and other dielectric crystals\cite{Debu2019,Zhou2021} has shown gate-voltage tunable hybrid plasmon-phonon polaritons (HPPPs). The HPPP modes provide anisotropic propagation of polaritons\cite{Maia2019} up to 2 times longer distances than hyperbolic PhPs\cite{Dai2015} with reduced damping\cite{Woessner2014}. However, there exists a wide spectral gap (400 cm$^{-1}$ to 1100 cm$^{-1}$) and uniaxial anisotropic h-BN crystal exhibits narrow Reststrahlen Bands (RBs)-- spectral region between LO and TO phonons of a particular polarization. Furthermore, the in-plane anisotropy of SERs for heterostructure of graphene with other hyperbolic materials like h-BN and LiNbO\textsubscript{3} is further restricted by their uniaxial nature\cite{Kumar2015,Debu2019,Zhou2021}. This challenge can be addressed via recently discovered natural vdW crystal\cite{Ma2018}, i.e. $\alpha$-MoO\textsubscript{3}, which exhibits biaxial hyperbolic anisotropy in the mid-infrared spectral region from 545 cm$^{-1}$ to 1006 cm$^{-1}$. Therefore, integration of graphene with $\alpha$-MoO\textsubscript{3} can provide highly anisotropic polarization dependent SERs over a broad spectral region without using any complex lithography techniques.\par

\begin{figure*}
    \includegraphics[width=\textwidth]{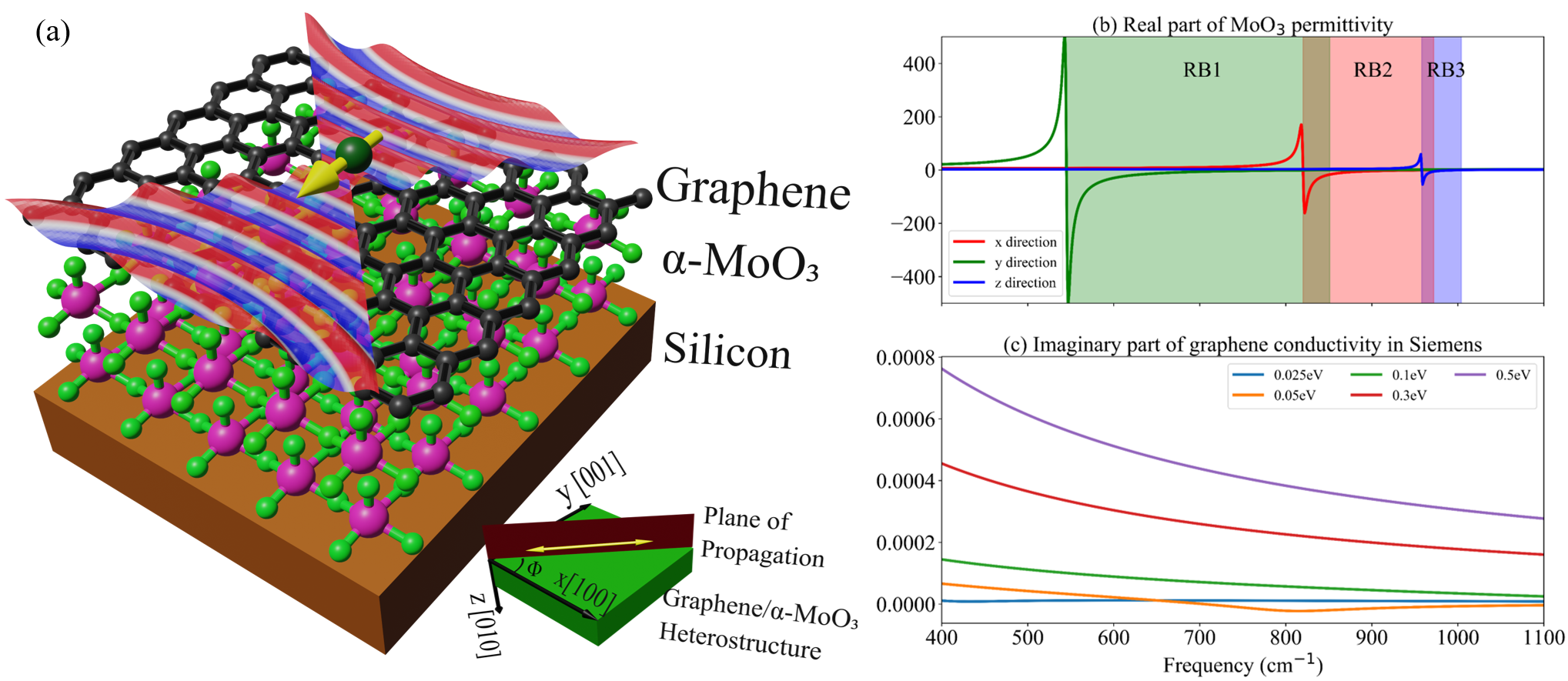}
    \caption{(a) Schematic representation of $\alpha$-MoO\textsubscript{3} and graphene heterostructure on silicon substrate exhibiting the hybrid plasmon-phonon polariton modes propagation at the surface. Inset: crystallographic direction of $\alpha$-MoO\textsubscript{3} and $\phi$-- angle between plane of propagation of light and x-axis of the $\alpha$-MoO\textsubscript{3}.{ (b) real component of dielectric permittivity of $\alpha$-MoO\textsubscript{3} along the principal directions. (c) imaginary component of 2D in-plane conductivities for varying chemical potential in graphene.}}
    \label{fig:1}
\end{figure*}

In this work, \textcolor{black}{we present a natural platform for tunable anisotropic HPPPs over a broad spectral region} (from 545 cm$^{-1}$ to 1100 cm$^{-1}$) via integration of graphene with biaxial hyperbolic vdW crystal, i.e. $\alpha$-MoO\textsubscript{3}, shown in Fig.~\ref{fig:1}. We investigate optical properties of air/graphene/$\alpha$-MoO\textsubscript{3}/Si heterostructure using the transfer matrix method (TMM)\cite{Schubert_transfer} as a function of chemical potential of graphene. Furthermore, we develop a framework by combining TMM with Green's dyadic function\cite{Lakhtakia1992,Novotny2006,gomez2015hyperbolic} to investigate anisotropic SERs from an electric dipole in the vicinity of the proposed heterostructure. \textcolor{black}{We found that the coupling of graphene SPPs with hyperbolic PhPs of $\alpha$-MoO\textsubscript{3} results in remarkably enhanced SERs over a broad mid-IR spectral region.} This investigation provides a lithography-free alternative for actively tunable anisotropic SERs via a sub-wavelength thick crystal of naturally biaxial hyperbolic materials. \textcolor{black}{As an example application, we present our results on gate tunable spontaneous valley coherence\cite{PhysRevLett.121.116102,PhysRevB.102.045416,Sohoni2020}, quantified by quantum interference, using this heterostructure platform.}

\section*{\label{RD}Results and Discussion}
\begin{figure*}
    \centering
    \includegraphics[width=\textwidth ]{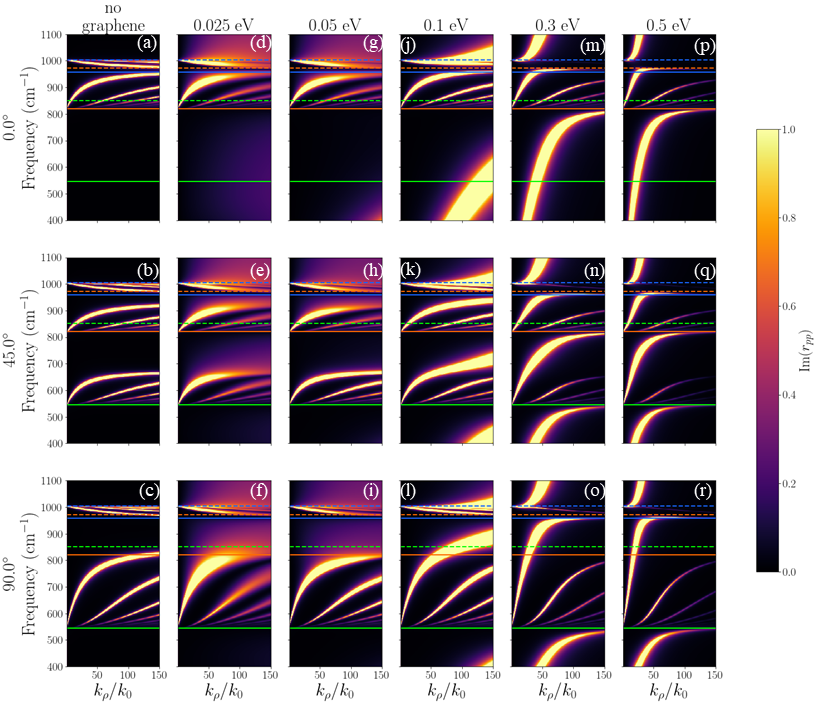}
    \caption{Dispersion of polaritonic modes in air/graphene/$\alpha$-MoO\textsubscript{3} (50nm)/Si heterostructure as a function of $\phi$ and chemical potential ($\mu$) of graphene. {Dotted lines indicate LO phonon frequencies, and solid lines indicate TO phonon frequencies along $y-$(green), $x-$(orange) and $z-$(blue) crystallographic directions of $\alpha$-MoO\textsubscript{3}.}}
    \label{fig:2}
\end{figure*}

\textbf{Biaxial hyperbolic phonon polaritons in $\alpha$-MoO\textsubscript{3}}. The $\alpha$-MoO\textsubscript{3} unit cell has lattice constants of around $0.396$ nm, $0.369$ nm, and $1.385$ nm in the $x$, $y$, and $z$ crystallographic directions\cite{Ma2018} as shown in Fig.~\ref{fig:1}{(a)}. The orthorhombic nature of the unit cell results in different sets of longitudinal optical (LO) and transverse optical (TO) phonons in the three crystallographic directions. The three sets of LO and TO phonons constitute three RBs in the mid-infrared spectral region, where $\alpha$-MoO\textsubscript{3} exhibits natural biaxial hyperbolicity. Spectral region for RB-1, RB-2, and RB-3 of $\alpha$-MoO\textsubscript{3} lie in the ranges of 545 cm$^{-1}$ - 850 cm$^{-1}$, 820 cm$^{-1}$ - 972 cm$^{-1}$, and 958 cm$^{-1}$ - 1006 cm$^{-1}$ respectively, where real part of dielectric permittivity is negative along $y$, $x$ and $z$ crystallographic directions, respectively, {as shown in Fig.~\ref{fig:1}(b)\cite{Zheng2019}. } Since $\alpha$-MoO\textsubscript{3} is a biaxial hyperbolic material, the dispersion of PhPs strongly depends on angle between plane of propagation of the polaritons and $x-$ crystallographic direction as shown in Fig.~\ref{fig:1}{(a)} and represented by $\phi$. We investigate the dispersion of PhPs { in an air/$\alpha$-MoO\textsubscript{3}/Si heterostructure} at $\phi = 0^\circ,\ 45^\circ$ and $90^\circ$ as shown in Fig.~\ref{fig:2}(a)--(c). Here, the thickness of $\alpha$-MoO\textsubscript{3} is chosen to be 50 nm, and the imaginary part of the Fresnel reflection coefficient for $p-$ polarized light is plotted in frequency-momentum space.\par

Fig.~\ref{fig:2}(a) reveals discrete modes in the RB-2 and RB-3 spectral regions for $\phi = 0^\circ$, which are attributed to the hyperbolic anisotropy of $\alpha$-MoO\textsubscript{3} resulting in the dispersion of fundamental and higher-order modes of hyperbolic PhPs. These modes demonstrate positive dispersion (i.e., v$_g > 0$) in RB-2 spectral region (Type-2 hyperbolicity) and negative dispersion (i.e., v$_g < 0$) in RB-3 spectral region (Type-1 hyperbolicity). Negative dispersion in the RB-3 spectral region is associated with the negative real part of dielectric permittivity in the out-of-plane crystallographic direction. We also observe similar PhP modes for $\phi=90^\circ$ in the RB-1 and RB-3 regions (Fig.~\ref{fig:2} (c)). \textcolor{black}{For $\phi = 45^\circ$ we observe Type-1 hyperbolic PhP modes in both RB1 and RB2 as the in-plane ($\epsilon_{11}$,$\epsilon_{12}$,$\epsilon_{21}$,$\epsilon_{22}$) entries of the rotated permittivity tensor depend on the permittivity along both $x$ and $y$ principal directions. However, these modes do not span the entire frequency range of either RB. Equation \ref{eq:1} suggests that the $\epsilon_{11}$ component of the rotated tensor is the most relevant in determining the extent of these PhP modes}.\par

\textbf{Hybridization of biaxial hyperbolic PhPs and SPPs.} We examine the hybridization of SPPs of graphene with hyperbolic PhPs of $\alpha$-MoO\textsubscript{3} via modulation of chemical potential ($\mu$) in graphene. We consider a heterostructure having the configuration: air/graphene/$\alpha$-MoO\textsubscript{3}/silicon as shown in Fig.~\ref{fig:1}(a) and explore the dispersion of polaritons, shown in the Fig.~\ref{fig:2}. {The conductivity of graphene for various values of chemical potential is shown in Fig.~\ref{fig:1}(c).} For $\mu$ = $0.025$ eV and $0.05$ eV, we observe smearing of hyperbolic PhP modes in all the RBs of $\alpha$-MoO\textsubscript{3} (Figs.~\ref{fig:2}(d)-(i)), however, there are no significant changes in the dispersion of hyperbolic PhPs in the plotted range of wave-vectors. This is due to interband transition in graphene upon excitation by a photon with an energy greater than $2\mu$ (around 403 cm$^{-1}$ and 806 cm$^{-1}$ for $\mu$ = $0.025$ eV and $0.05$ eV, respectively).At $\mu$ = $0.1$ eV in Fig.~\ref{fig:2}(j), a broad mode is observed above the RB-3 spectral region and below the RB-2 spectral region. This mode corresponds to the SPPs of graphene, since as the chemical potential increases, the graphene plasmon dispersion moves closer to the light cone\cite{Kumar2015}. \textcolor{black}{Due to the inherently lossy nature of the SPP mode, it is broader than the phonon modes.} Furthermore, in Figs.~\ref{fig:2}(k)-(l), a variation is observed in the dispersion of the fundamental mode of hyperbolic PhPs in the RB-1 spectral region. This is ascribed to the hybridization of hyperbolic PhPs of $\alpha$-MoO\textsubscript{3} with SPPs of graphene forming HPPPs. Moreover, for $\mu$ = $0.3$ eV and $0.5$ eV, the dispersion of hyperbolic PhPs in RB-1 and RB-2 spectral region become steeper compared to the no graphene case, which is a strong indication of hybridization of hyperbolic PhPs of $\alpha$-MoO\textsubscript{3} with SPPs of graphene (Figs.~\ref{fig:2}(m)-(r)). \textcolor{black}{Additionally, we observe that the dispersion curve of SPPs becomes nearly horizontal as it approaches the TO phonon frequencies of $\alpha$-MoO\textsubscript{3}.} This behaviour is due to the avoided-crossing\cite{Kumar2015,Debu2019}, which suggests a coupling between surface plasmons of graphene and TO phonons of $\alpha$-MoO\textsubscript{3} and further substantiates the formation of HPPPs. \textcolor{black}{Dispersion curves in Figs.~\ref{fig:2}(o) and \ref{fig:2}(r) demonstrate that these hybrid modes can propagate outside the RB-1 and RB-2 spectral region respectively, providing additional operational bandwidth compared to HPhPs of $\alpha$-MoO\textsubscript{3} and relatively less losses compared to the SPPs of graphene.} 

Apart from fundamental modes of hyperbolic PhPs in RB-1 and RB-2, variation in the dispersion of higher-order PhP modes is observed upon integration of graphene. For a particular mode index ({k}$_\rho$/k$_0$), a blue-shift is observed in the frequencies of higher-order modes with increasing chemical potential of graphene, which we will discuss in the sections below. Similar to the observed trend in RB-1 and RB-2 spectral regions, hyperbolic PhP modes in the RB-3 spectral region are strongly dominated by graphene's interband transitions at lower $\mu$ values ($0.025$eV and $0.05$eV) and graphene's SPPs at higher $\mu$ values. At $\mu = 0.3$eV and $0.5$eV, mode repulsion is observed near the out-of-plane TO phonon frequency of $\alpha$-MoO\textsubscript{3} which indicates the coupling of SPPs with these TO phonons. A discussion of the effect of thickness of  $\alpha$-MoO\textsubscript{3} on dispersion of HPPPs is presented in the Sec.~S2 of supplementary information.

\begin{figure}
    \centering
    \includegraphics[scale=0.65]{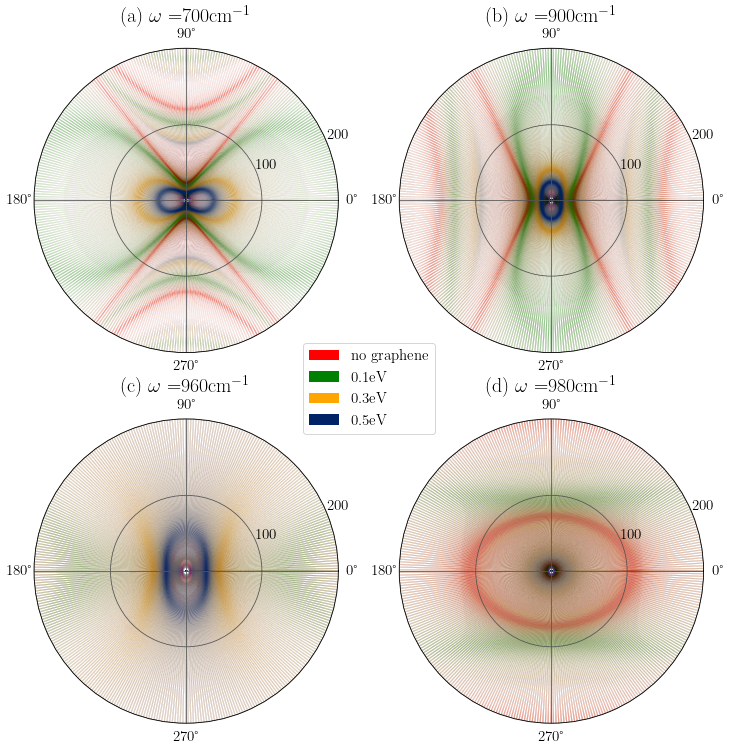}
    \caption{Polar plots of HPPP modes in 50nm thin $\alpha$-MoO\textsubscript{3} and graphene heterostructure as a function of graphene's chemical potential (0.1, 0.3, 0.5 eV): HPPP fundamental and first-order modes are represented by imaginary part of Fresnel's reflection coefficient for $p-$ polarized light. Radial and azimuthal directions correspond to mode index ($k_\rho/k_0$) and $\phi$, respectively. Effect of graphene's chemical potential on HPPP modes is demonstrated by green, orange and blue color for $\mu$ = $0.1$eV, $0.3$eV, and $0.5$eV respectively. The fringes at the edges of the polar plots are the artefact of the scatter type plotting scheme.}
    \label{fig:3}
\end{figure}

\begin{figure*}
    \centering
    \includegraphics[width=\textwidth]{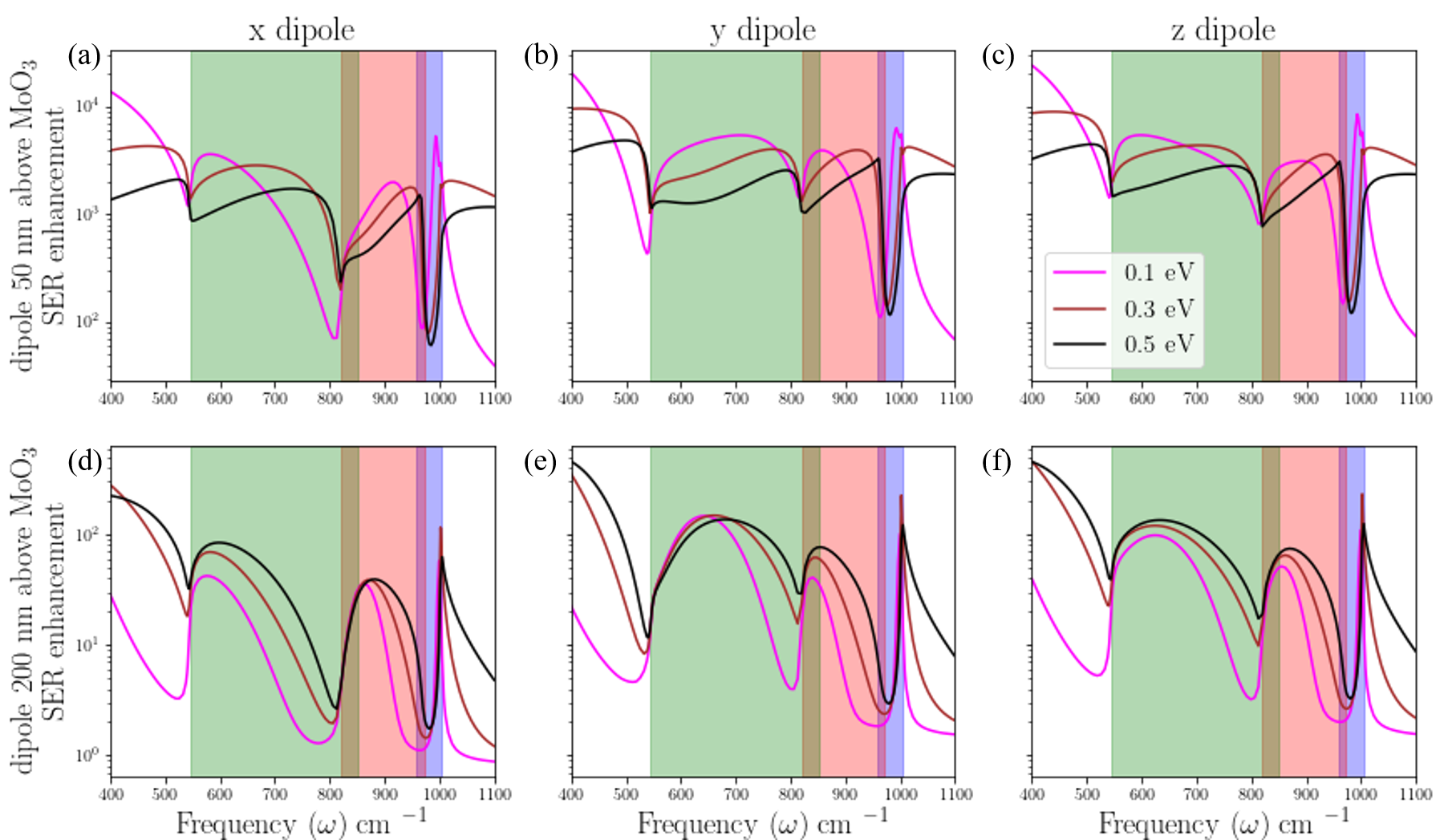}
    \caption{Anisotropic spontaneous emission rates for dipole placed above air/graphene/$\alpha$-MoO\textsubscript{3} (50 nm)/Si heterostructure : (a)-(c) represent SER enhancements at different $\mu$ values for $x-$, $y-$ and $z-$ polarized electric dipole, respectively, at $z_0 = 50$nm. (d)-(f) represent SER enhancements at different $\mu$ values for $x-$, $y-$ and $z-$ polarized electric dipole, respectively, at $z_0 = 200$nm. The green, red and blue shaded regions represent the RB-1, RB-2, and RB-3 spectral regions of $\alpha$-MoO\textsubscript{3}, respectively}
    \label{fig:4}
\end{figure*}

Hybridization of SPPs and hyperbolic PhPs is further investigated via iso-frequency surfaces for these polaritons, plotted as a function of $\phi$, $\mu$, and $k_\rho/k_0$. We consider $\omega$ = 700 cm$^{-1}$, 900 cm$^{-1}$, 960 cm$^{-1}$ and 980 cm$^{-1}$, which are representative of frequencies in the RB-1, RB-2, transition region between RB-2 and RB-3, and RB-3 spectral regions in Fig.~\ref{fig:3}. At $\omega$ = 700 cm$^{-1}$ in Fig.~\ref{fig:3}(a), we observe that the fundamental and higher-order hyperbolic PhP modes of $\alpha$-MoO\textsubscript{3} appear as hyperbolae aligned in the $\phi = 90^\circ$ direction (i.e. $y$- direction). The iso-frequency surfaces of the heterostructure of $\alpha$-MoO\textsubscript{3} and graphene with $\mu = 0.1$eV, $0.3$eV, $0.5$eV appear as ellipses squeezed along $\phi = 90^\circ$ due to the hybridization of hyperbolic PhPs of $\alpha$-MoO\textsubscript{3} with the isotropic SPPs of graphene along $y-$ direction. Furthermore, we observe a shift towards lower mode index in the first-order mode of hyperbolic PhPs with increasing chemical potential of graphene compared to that of no graphene (see Fig.~S5 of supplementary information). Fig.~\ref{fig:3}a confirms that the higher order PhP modes are affected by graphene integration with $\alpha$-MoO\textsubscript{3}. For $\mu = 0.3$eV and $0.5$eV, the SPP wave vector decreases resulting in smaller ellipses compared to the $\mu = 0.1$eV case. Similarly, at $\omega$ = 900 cm$^{-1}$ in Fig.~\ref{fig:3}(b), the hyperbola is found aligned along $\phi = 0^\circ$ ($x-$ direction) and first-order mode inside hyperbola varies with increasing chemical potential. These results confirm the presence of tunable hybridization of fundamental and higher-order PhP modes with SPPs via chemical potential of graphene. Polar plots for other chemical potentials at $\omega$ = 700 cm$^{-1}$ and 900 cm$^{-1}$ are presented in the supplementary information. At $\omega = 960$cm$^{-1}$, we do not observe hyperbolic PhP modes for the no graphene case in Fig.~\ref{fig:3}(c). This is associated with the inherent dielectric response of $\alpha$-MoO\textsubscript{3} in the transition from RB-2 to RB-3 spectral region. However, graphene SPPs dominate the formation of HPPPs for $\mu= 0.3$eV and $0.5$eV as shown by the squeezed ellipse in the Fig.~\ref{fig:3}(c) and Fig.~S6 of the supplementary information. Moreover, at $\omega$ = 980 cm$^{-1}$, we observe the transition of hyperbolic contour into an elliptical contour for the no graphene case as shown in Fig.~\ref{fig:3}(d). For $\mu = 0.1$ eV, we observe that the elliptical contour is elongated along $x-$ crystalline direction, which is attributed to mode repulsion between HPPP modes in the RB-2 and RB-3 spectral regions for $\phi=0^{\circ}$. As a result, we observe higher mode index at 980 cm$^{-1}$ along $x-$ direction. On the other hand, HPPP modes in the RB-2 spectral region do not exist for $\phi=90^{\circ}$ and consequently, we have a lower mode index at 980 cm$^{-1}$ along $y-$ direction as compared to the $x-$ crystalline direction (see Fig.~S6 of the supplementary information). Hence the elliptical contour for $\mu = 0.1$eV is elongated along $x-$ crystal direction. Lastly, at $\mu = 0.3$eV and $0.5$eV, we observe a feeble elliptical contour in Fig.~S4 of the supplementary information (which are absent in Fig.~\ref{fig:3}(d) is due to relatively greater strength of contour for no graphene case). This can be attributed to type-1 HPPPs in the RB-3 spectral region. After integration of graphene with $\mu = 0.3$eV and $0.5$eV, the HPPP modes in the RB-3 spectral region move upward (or towards higher mode index) and create a spectral gap around 980 cm$^{-1}$, (see Figs.~S4 and S6 of supplementary information) resulting in a feeble elliptical contour.\par

{In order to understand the origin of this active tunability and the angle dependence of these hybrid modes, we derive an analytical expression for these dispersion relations under the quasistatic approximation ($k_\rho \gg k_0$). The expression for the dispersion curve then is given by:
\begin{align}
    \frac{k_\rho}{k_0}(\omega)=& \frac{\psi}{k_0d}\biggl[\tan^{-1}\left(\frac{\psi (\epsilon_1 + i\left(\frac{k_\rho}{k_0}\right)\sigma Z_0)}{\epsilon_z}\right)\nonumber \\ 
    & + \tan^{-1}\left(\frac{\epsilon_3\psi}{\epsilon_z}\right) + n\pi\biggr] \label{eq:1}
\end{align}
Here, $\sigma$ is the conductivity of graphene, $Z_0 = \sqrt{{\mu_0}/{\epsilon_0}}$ is the impedance of free space, $\epsilon_z$ is the z-direction (out of plane) permittivity of $\alpha$-MoO\textsubscript{3} and $\psi = i\sqrt{{\epsilon_z}/{(\epsilon_x\cos^2\phi + \epsilon_y\sin^2\phi)}}$ and $n$ is an integer. $\epsilon_1 = 1$ and $\epsilon_3=2.5$ are taken as the permittivities of the air and Si substrate, respectively. This expression can be derived by imposing the condition that the total of phase shift due to reflection at the two interfaces of the $\alpha$-MoO\textsubscript{3} slab plus the propagation phase shift should be an integral multiple of $2\pi$ (see supplementry information). The variation of the graphene Fermi level through a gate voltage modifies $\sigma$ which affects the reflection phase shift at that interface, thereby enabling the tunability of the hybrid modes. It must be noted that all the results presented in this work are full wave calculations without using the quasi-static approximation.}

\begin{figure*}
    \centering
    \includegraphics[width=\textwidth]{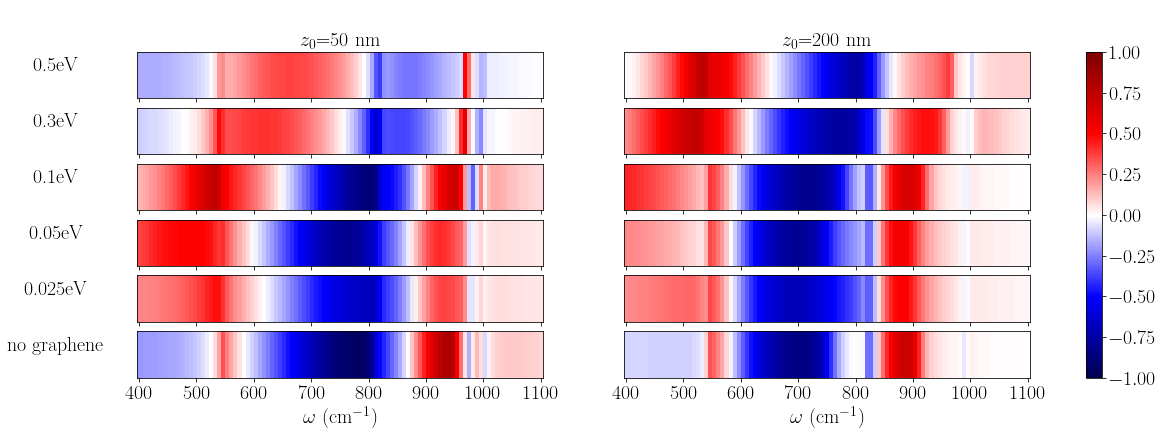}
    \caption{Quantum interference ($Q$) for air/graphene/MoO\textsubscript{3}/Si heterostructure with MoO\textsubscript{3} thickness of 50nm, plotted as a function of frequency, for two different dipole heights of $z_0 =50$ nm and $200$ nm, respectively.}
    \label{fig:5}
\end{figure*}

\textbf{Anisotropic spontaneous emission via biaxial hybrid polaritons.} Next, we explore the anisotropic SER enhancement from an electric dipole placed in the vicinity of graphene/$\alpha$-MoO\textsubscript{3}/Si heterostructure, as shown in Fig.~\ref{fig:1}. To the best of our knowledge, anisotropic SERs for natural biaxial hyperbolic material have not been reported in the literature so far. Since $\alpha$-MoO\textsubscript{3} exhibits in-plane hyperbolicity, we examine the anisotropic SERs by considering the polarization of the electric dipole along $x-$, $y-$, and $z-$ directions. Distance of the electric dipole from the surface of the heterostructure (represented by $z_0$) is taken to be 50 nm and 200 nm, whereas $\mu$ is varied as $0.1$ eV, $0.3$ eV, and $0.5$ eV, as shown in Fig.~\ref{fig:4}.  For comparison, we have also evaluated the anisotropic SERs through $\alpha$-MoO\textsubscript{3} without graphene and for lower $\mu$ values, shown in the Sec.~S4 of the supplementary information. Integration of graphene with $\alpha$-MoO\textsubscript{3} crystals not only provides tunable spontaneous emission but also enhances the SER to a much broader operational bandwidth.\par 
At $z_0 = 50$ nm, the order of magnitude of SER is found to be more than 10$^3$ inside the RB spectral regions for $x$, $y$ and $z$ oriented dipoles as shown in Fig.~\ref{fig:4}(a)--(c) respectively. Anisotropic SER spectra are dominated by graphene SPP modes outside the RBs of $\alpha$-MoO\textsubscript{3}. The difference between SER spectra in the RB spectral regions with and without integration of graphene and $\alpha$-MoO\textsubscript{3} is associated with the formation of HPPPs, which provides a larger operational bandwidth. Variation of SER spectra in the RBs as a function of chemical potential is due to the change in the  dispersion of HPPP modes with chemical potential, as discussed above. Spectral dips in SERs spectra are attributed to the spectral gap due to the absence of strongly confined SPP modes and hyperbolic PhP modes in the avoided crossing regions of the dispersion. \textcolor{black}{Within such an avoided crossing, the dipole cannot couple to any mode effectively and hence spontaneous emission is suppressed.} Here, we see that the line-widths of spectral dips decrease with increasing chemical potential, which is due to the fact that the graphene plasmon mode index decreases upon increase of chemical potential\cite{Kumar2015}, decreasing the strength of the avoided crossing at larger mode indices just below the RBs of $\alpha$-MoO\textsubscript{3}. Moreover, a blue-shift in the spectral dip is observed from 960cm$^{-1}$ to 980cm$^{-1}$ when $\mu$ increases from $0.1$eV to $0.3$eV. As described above with Figs.~\ref{fig:3}(c) and \ref{fig:3}(d), this is due to the shift of the RB-3 HPPPs towards higher mode index as the graphene Fermi level increases -- this trend is opposite to that of the HPPPs in the other bands.\par
At $z_0 = 200$nm, (shown in Fig. \ref{fig:4}(d)-(f)), we observe a decrease in the SER by one order of magnitude due to the evanescent nature of SPPs and HPPPs, resulting in reduced interaction of polaritonic modes with the electric dipole and hence reduction of the SERs. The dips in the SER spectra are due to the spectral gap of polariton modes, as discussed above. Reduction in the line-width of spectral dips with increasing chemical potential is found as expected. Below RB-1 spectral region, the SER for $\mu= 0.3$eV and $0.5$eV is higher in comparison with smaller $\mu$ values. Since higher chemical potential increases the imaginary part of the graphene conductivity, it reduces the confinement of SPPs. This can be seen from dispersion of SPPs in Fig.~\ref{fig:2}, where one can observe SPP modes at higher $\mu$ values just below the RB-1 spectral region move to smaller $k_\rho$ values. As a result, SPP modes for higher chemical potential have an evanescent tail at a larger distance away from the surface compared to the SPP modes for smaller chemical potential and hence the SER can be enhanced. A mathematical justification can further be provided based on the envelope $k_{\rho}^2e^{{-2 k_{\rho}z_0}}$ which peaks at $k_{\rho}\sim 1/z_0$. So depending on whether the mode is resonant at $k_{\rho}\sim 1/z_0$, one can observe an increase in the SER. The variation of the anisotropic SER spectra as a function of the thickness of $\alpha$-MoO\textsubscript{3} are discussed in the Sec.~S4 of the supplementary information. \textcolor{black}{Similarly, spontaneous emission spectra for the air/graphene/Si heterostructure is discussed in Sec.~S5 of the supplementary information.} These SER spectra substantiate our claim that one can observe a remarkable spontaneous emission enhancement for an electric dipole in the vicinity of $\alpha$-MoO\textsubscript{3} integrated with graphene over a broad spectral range.\par

\textcolor{black}{\textbf{Application to valley quantum interference:} We demonstrate the power of such a tunable anisotropic SER for application in spontaneous valley coherence\cite{PhysRevB.102.045416}. Quantum coherence in solid state systems is important not only from a fundamental point of view but is a key ingredient for realizing the promised applications in quantum technologies like quantum computing\cite{knill2001scheme}, quantum key distribution\cite{gisin2002quantum} and quantum metrology\cite{Giovannetti2011}. In the infrared range, gapped Dirac systems\cite{PhysRevLett.114.256601,PhysRevB.93.041413} such as biased bilayer graphene\cite{doi:10.1126/science.aam9175,Henriques2022} and similar systems could serve as an important excitonic building block for such technologies. The electronic bandstructure of these systems consists of two degenerate valleys $K$ and $K'$, in which the optical selection rule is sensitive to the polarization ($\sigma^{\pm}$) of the photon. In order to harness this ``valley degree of freedom", it is important to be able to actively control the coherence between excitons in the two valleys. While this coherence has been demonstrated in the stimulated regime\cite{Jones2013}, where the valley excitons inherit coherence from the exciting laser, a spontaneous route to such coherence generation would greatly simplify the device architecture and further, as we show below, enable electrostatic control of valley coherence, which could be easily implemented on a chip and provide a low-cost, facile fabrication and chip-scale alternatives for building relevant technologies.}\par

\textcolor{black}{Valley coherence can be quantified experimentally using the degree of linear polarization or DoLP of the emission, where $\text{DoLP}=(I_X-I_Y)/(I_X+I_Y)$, $I_X$ and $I_Y$ being the intensities of the two linearly polarized emissions. It can easily be shown that in the presence of a weak pump, the DoLP is approximately equal to the quantum interference $Q$, which is defined as\cite{PhysRevB.102.045416,PhysRevLett.121.116102}:}
\begin{equation}
    Q = \frac{\gamma_x - \gamma_y}{\gamma_x+\gamma_y}
\end{equation}
where $\gamma_x$ and $\gamma_y$ are the Purcell factors for $x$ and $y$ polarized dipoles, respectively. The geometrical configuration for this experiment would involve a gapped Dirac system such as a bilayer graphene system placed a certain distance away from the monolayer graphene/$\alpha$-MoO\textsubscript{3} system. The exciton of this gapped Dirac system will be represented by the dipole. The quantum interference for this heterostructure with a 50 nm thick $\alpha$-MoO\textsubscript{3} layer is plotted as a function of frequency for various values of chemical potential $\mu$ of graphene in Fig.~\ref{fig:5}. Since $\alpha$-MoO\textsubscript{3} is a naturally biaxial material, we observe large magnitude of anisotropy in RB-1 and RB-2 spectral regions, where dipole is aligned along the $y-$ and $x-$ crystallographic axes, respectively, and can couple more strongly to PhP modes as compared to the dipole polarized orthogonal to the metallic direction of the hyperbolic crystal.\par

Integration of graphene at chemical potentials of 0.025eV and 0.05eV suppresses the quantum interference, as for the chosen dipole distances, the emitter does not couple efficiently to the HPPPs as they occur at very large wave vectors (as seen in Fig.~\ref{fig:2}) following the relation $k_{\rho}\sim 1/z_0$ and the graphene itself is isotropic. However, when the chemical potential is changed to 0.1 eV, 0.3 eV, and 0.5 eV, we begin to see significant differences in the quantum interference, due to coupling with the HPPP modes. For $z_0 = 50$ nm, we see large positive degree of anistropy just below 545 cm$^{-1}$. This is explained by the avoided crossing between the graphene plasmon and phonon polaritons in RB-1 region, when the plane of propagation is along y-z plane (i.e. $\phi= 90^\circ$). However, this phenomenon does not occur in the x-z plane (i.e. $\phi= 0^\circ$) for frequencies below RB-1. Thus, the $y$-dipole sees an avoided crossing unlike the $x$ dipole, and we get large positive anisotropy. Furthermore, a shift in the frequency band of strong anisotropy is observed with increasing chemical potential of graphene from 0.025eV to 0.5eV inside RB-1 and RB-2 spectral region which is due to active tuning of HPPP modes, as discussed above.

Moreover, for $z_0 = 200$ nm, we observe a similar phenomenon at $\mu = 0.1$eV$-0.5$eV. As we increase $\mu$, the SPPs mode moves to a lower $k_\rho/k_0$. Since large $k_\rho/k_0$ modes are tightly confined, they remain unavailable to dipoles placed far away from hetero-structure. Hence, this lower wave-vector modes at large $\mu$ are more prominent for a dipole at $z_0 =$ 200nm as can be observed in Fig.~\ref{fig:5}(b). We also observe that the frequency corresponding to large negative anisotropy shifts at the upper end of the RB-1 which, as described above, is due to the direction dependent avoided crossing at the lower end of the RB-2 for one emitter polarization as opposed to coupling with the isotropic graphene SPP outside the RB-1 for the other polarization. Thus, we see a large value of negative anisotropy just below RB-2, instead of a nearly zero value after RB-1.

We also see that the large negative anisotropy for most of the RB-1 region in the no graphene case is replaced by a small positive anisotropy when 0.3 eV or 0.5 eV graphene is introduced. This can be understood from Fig.~\ref{fig:3} where we observe that while there are no available modes in $x-$ direction without graphene (hyperbola), the lemniscate shape of the fundamental HPPP mode with $\mu=$ 0.3eV and 0.5eV for graphene provides available modes to both dipole orientations. We also see a corresponding reverse phenomenon in RB-2 region in Fig.~\ref{fig:5}, which has a similar explanation. Thus, the proposed heterostructure provides a convenient way of electrostatically tuning anisotropy at a given frequency.

\section*{\label{conclusion} Conclusion}
In conclusion, we have investigated the integration of graphene with recently discovered biaxial hyperbolic vdW crystal of $\alpha$-MoO\textsubscript{3} for applications in quantum photonics. We found a unique direction dependent hybridization of surface plasmon-polaritons of graphene with fundamental and higher-order modes of hyperbolic phonon polaritons in $\alpha$-MoO\textsubscript{3} crystals forming hybrid plasmon phonon polaritons. We show that these HPPPs can be actively tuned via electrostatic gating of graphene and can exist outside the RBs of $\alpha$-MoO\textsubscript{3}, providing enhanced operational bandwidth. We have found significantly large anisotropic SER enhancements of more than 10$^3$ in the vicinity of the proposed heterostructure with an operational bandwidth from 400 cm$^{-1}$ to 1100 cm$^{-1}$ and report large and actively tunable quantum interference for SER.\par 
Compared to uniaxial hyperbolic materials, the tunable biaxial hyperbolic heterostructure proposed in this work has unique advantages in terms of strong in-plane anisotropy leading to precise control of energy channeling direction. The current platform also overcomes the issue of complex lithographic fabrication for schemes implementing biaxial hyperbolicity through metasurfaces\cite{GomezDiaz2016} since our platform relies on the intrinsic phononic and plasmonic response of the constituent materials of the heterostructure.\textcolor{black}{This heterostructure also provides an interesting case for the study of exceptional point physics \cite{park2020graphene}.} The tunability of the isofrequency surfaces of the HPPP modes could find application in tunable radiative heat flow control including radiative cooling and exotic phenomenon such as tunable super-Planckian near-field thermal emission\cite{Wang2020H,Guo2013,Guo2012,NFRHT_MoO3_gra}. The broadband and strong light-matter interaction enabled by these heterostructure could be a powerful tool for vibrational spectroscopy particularly for biosensing, where electrostatic gating could enable not only high sensitivity and tunable spectral selectivity\cite{Rodrigo2015,Hu2019s,Oh2021} but also serve as a unique probe of orientation of the molecule due to the in-plane hyperbolic nature of the HPPPs in our heterostructures. Our work opens new avenues for tunable in-plane hyperbolic polaritonics for other exotic applications in quantum interference\cite{PhysRevLett.121.116102,PhysRevB.102.045416,Sohoni2020,PhysRevB.95.075412,PhysRevA.101.013837,Karanikolas2018}, gate tunable planar refraction devices using polaritons\cite{Duan2021}, emission pattern engineering\cite{Zhang2019}, quantum field entanglement\cite{PhysRevA.96.043865}, engineering interaction between multiple quantum emitters\cite{Newman2018,Cortes2018,Cortes2017} and Kerr nonlinearity enhancement\cite{Chen2015}.\par

\emph{Note:} After the submission of our manuscript, these recent related works were brought to our notice\cite{Yadav2021,perez2021active}.

\section{\label{methods} Methods}
\noindent\textit{\textbf{Model for calculation of polariton dispersion:}} We used the standard expression for complex valued 2D graphene conductivity derived from the local limit of the random phase approximation\cite{Debu2019,Graphene_plasmonics_RPA}

\begin{equation}
\begin{split}
    \sigma(\Omega,\mu) &= \frac{ie^2k_BT}{\pi\hbar^2(\Omega + i/\tau)}\left[\frac{\mu}{k_BT}+2\ln\left(exp\left(-\frac{\mu}{k_BT}\right)+1\right)\right] \\
    &\quad{} +\frac{ie^2}{4\pi\hbar}\ln\left(\frac{2\mu - \hbar(\Omega+i/\tau)}{2\mu + \hbar(\Omega+i/\tau)}\right)
\end{split}
\label{Eq:3}
\end{equation}

Here, the first-term represents the intraband contribution to graphene conductivity and the second term is the interband contribution. $e,k_B$ and $\hbar$ refer to the charge of electron, Boltzmann's constant and reduced Planck constant respectively. We considered the value of temperature as $T = 300K$ and relaxation time $\tau = 10^{-13} s$ which matches well with experimental results \cite{Graphene_plasmonics_RPA}. Here $\mu$ refers to the and the chemical potential of graphene chosen (in eV). $\Omega = 200\pi c \omega$ is the angular frequency in rad/s (where $\omega$ is the frequency in cm$^{-1}$). This conductivity expression has been plotted for various chemical potentials $(\mu)$ in the supplementary document (Figure S1). Dielectric permittivity of graphene is derived using the Drude model to be \cite{graphene_permittivity}
\begin{equation}
    \epsilon_{gr} = 1 + \frac{i\sigma(\omega,\mu)}{\epsilon_0 \omega d}
    \label{Eq:S2}
\end{equation}
where $\epsilon_{gr}$ is the relative permittivity of graphene in the in-plane directions, $\epsilon_0$ is the permittivity of free space, and the thickness of the monolayer graphene ($d$) is taken 0.34nm in accordance with experimental results \cite{graphene_thickness_expt}. Since the out-of-plane conductivity of graphene is zero, the relative permittivity of graphene in the out-of-plane direction is unity.\par

The permittivity of $\alpha$-MoO\textsubscript{3} in mid-IR spectral region is given by the Lorentz oscillator model. Frequency-dependent dielectric permittivity along each crystal axis is written as 
\begin{equation}
    \epsilon_i(\omega) = \epsilon_{\infty,i}\left(1 + \frac{\omega_{LO,i}^2 - \omega_{TO,i}^2}{\omega_{TO,i}^2 - \omega^2 - i\gamma_{i}\omega}\right)
    \label{Eq:S3}
\end{equation}
We used the following values for the parameters $\epsilon_\infty,\omega_{LO},\omega_{TO}$ and $\gamma$ along each principal axis, as proposed in \cite{Zheng2019}.

\begin{center}
    \begin{tabular}{|c|c|c|c|}
         \hline
         &     $x$     &     $y$     &     $z$       \\
         \hline
         $\epsilon_\infty$& 4.0 & 5.2 & 2.4\\
         \hline
         $\omega_{LO}$ (cm$^{-1}$) & 972 & 851 & 1004\\
         \hline
         $\omega_{TO}$ (cm$^{-1}$) & 820 & 545 & 958\\
         \hline
         $\gamma$ (cm$^{-1}$) & 4 & 4 & 2\\
         \hline
    
    \end{tabular}
\end{center}
The respective expressions for permittivities along principal directions are plotted in the supplementary information (Figure S1).

Permittivity tensor for any particular plane of propagation (i.e. for any $\phi$) is obtained from $\epsilon(\phi,\omega) = R(\phi)\epsilon_{principal}(\omega) R^T(\phi)$ where

\begin{equation}
R(\phi) = \begin{pmatrix}\cos\phi & -\sin\phi & 0 \\ \sin\phi &\cos\phi &0\\ 0&0&1\end{pmatrix}   
\label{Eq:S4}
\end{equation}
is the rotation matrix for transformation of coordinate system, and $\epsilon_{principal}(\omega)$ is the permittivity tensor in principal axis coordinate system \cite{Schubert_transfer}. 
{This leads to the final expression - 
$$\epsilon(\phi,\omega) = \begin{pmatrix}\epsilon_x\cos^2\phi + \epsilon_y\sin^2\phi&\sin\phi\cos\phi(\epsilon_y-\epsilon_x)&0\\ \sin\phi\cos\phi(\epsilon_y-\epsilon_x)&\epsilon_x\sin^2\phi + \epsilon_y\cos^2\phi&0\\0&0&\epsilon_z\end{pmatrix}$$}\\
Fresnel's reflection coefficient for $p-$ polarized light is evaluated from well-known $4\times4$ transfer matrix method \cite{Schubert_transfer}.\par
\noindent\textit{\textbf{Calculation of Purcell Factor:}}
Purcell factor for an emitter placed at a height $z_0$ above a surface in terms of the Green's Dyadic function can be written as \cite{Novotny2006,Lakhtakia1992,gomez2015hyperbolic}
\textcolor{black}{\begin{equation}
    \gamma = \frac{P}{P_0} = 1 + \frac{6\pi}{k_0}\Vec{\mu}\cdot Im(G)\cdot \Vec{\mu}
    \label{Eq:S5}
\end{equation}}
where $\Vec{\mu}$ is a unit vector along the direction of the dipole, $G=G(\vec{r_0},\vec{r_0})$ is the scattering part of the Green's dyadic function at the position of the dipole, and $k_0 = \frac{\omega}{c}$. \textcolor{black}{Here, $P$ and $P_0$ are the spontaneous emission rates of a dipole placed above the heterostructure and in free space, respectively.}

From \cite{Lakhtakia1992}\cite{gomez2015hyperbolic} we can express the Green's tensor as - 
\begin{multline}
G = \frac{i}{8\pi^2}\iint (r_{ss}M_{ss}+r_{sp}M_{sp}+r_{ps}M_{ps}+r_{pp}M_{pp})e^{2ik_zz_0}\\
dk_xdk_y
\label{Eq:8}
\end{multline}
Where the integral is taken over all space and `$M$' matrices are given by:
\begin{align*}
    M_{ss}&=\frac{1}{k_zk_\rho^2}\begin{pmatrix}k_y^2&-k_xk_y&0\\-k_xk_y&k_x^2&0\\0&0&0\end{pmatrix} \\
    M_{sp}&=\frac{1}{k_0k_\rho^2}\begin{pmatrix}-k_xk_y & -k_y^2 & -\frac{k_yk_\rho^2}{k_z}\\k_x^2 & k_xk_y & -\frac{k_xk_\rho^2}{k_z}\\0&0&0 \end{pmatrix}\\
    M_{ps}&=\frac{1}{k_0k_\rho^2}\begin{pmatrix}k_xk_y & -k_x^2 & 0 \\k_y^2 & -k_xk_y & 0\\-\frac{k_yk_\rho^2}{k_z}&\frac{k_xk_\rho^2}{k_z}&0 \end{pmatrix}\\
    M_{pp}&=\frac{k_z}{k_0^2k_\rho^2}\begin{pmatrix}-k_x^2 & -k_xk_y & -\frac{k_xk_\rho^2}{k_z}\\-k_xk_y & -k_y^2 & -\frac{k_yk_\rho^2}{k_z}\\\frac{k_xk_\rho^2}{k_z}&\frac{k_yk_\rho^2}{k_z} & \frac{k_\rho^4}{k_z^2} \end{pmatrix}
\end{align*}
From here, we can recast the integral in polar coordinates and calculate the spontaneous emission rate enhancements.

\bibliography{main}

\end{document}